\documentstyle[12pt,latexsym]{article}
\def\beqa{\begin{eqnarray}}
\def\eeqa{\end{eqnarray}}
\def\beq{\begin{equation}}
\def\eeq{\end{equation}}
\newcommand{\ba}{\begin{array}}
\newcommand{\ea}{\end{array}}
\newcommand{\ft}[2]{{\textstyle{#1\over#2}}}
\newcommand{\fft}[2]{{#1\over#2}}

\def\text#1{\hbox{#1}}

\begin{document}

%%%%%%%%%%%%%%%%%%%%

\hbox{}
\begin{flushright}
\vspace{-2cm}
CAMS/00-08\\
UM-TH-00-25\\
hep-th/0010025\\
\end{flushright}

\vspace{10pt}

\begin{center}
{\Large {\bf Multi-centered black holes in}}
\bigskip

{\Large {\bf gauged $D=5$ supergravity}}

\vspace{20pt}

James T. Liu$\,^{\dagger }$, W. A. Sabra$\,^{\ddagger }$

\vspace{10pt} {$\, ^{\dagger}$ {\it Randall Laboratory, Department of
Physics, University of Michigan,\\[0pt]
Ann Arbor, MI 48109--1120}}

\vspace{5pt}

{\leavevmode\kern-2em$\,^{\ddagger}$ {\it Center for Advanced Mathematical
Sciences (CAMS) and Physics Department,\kern-2em\\[0pt]
American University of Beirut, Lebanon}}

\vspace{30pt}

ABSTRACT
\end{center}

One of the important consequences of the no-force condition for BPS states
is the existence of stable static multi-center solutions, at least in
ungauged supergravities.  This observation has been at the heart of many
developments in brane physics, including the construction of intersecting
branes and reduced symmetry D-brane configurations corresponding to the
Coulomb branch of the gauge theory.  However the search for multi-center
solutions to gauged supergravities has proven rather elusive.  Because of the
background curvature, it appears such solutions cannot be static.
Nevertheless even allowing for time dependence, general multi-center
solutions to gauged supergravity have yet to be constructed.  In this letter
we investigate the construction of such solutions for the case of
$D=5$, $N=2$ gauged supergravity coupled to an arbitrary number of vector
multiplets.  Formally, we find a family of time dependent multi-center
black hole solutions which are easily generalized to the case of AdS
supergravities in general dimensions.  While these are not true
solutions, as they have a complex metric and gauge potential, they may
be related to a Wick rotated theory or to a theory where the
coupling is taken to be imaginary.  These solutions thus provide a
partial realization of true multi-center black-holes in gauged
supergravities.

\newpage

%%%%%%%%%%%%%%%%%%%%

\section{Introduction}

There has been much interest in the study of black hole solutions of
ungauged and gauged supergravity theory in recent years. This is to a large
extent motivated by the study of duality symmetries and their relation to
the non-perturbative limits of string theory and M theory.  In addition
to providing the foundation for a microscopic understanding of black hole
physics, such explicit solutions play an important role in the
conjectured AdS/CFT correspondence.  Here, the anti-de Sitter geometry
arises as the vacuum of gauged supergravities in various dimensions.
Thus black holes in gauged supergravities are especially relevant for
investigations of AdS/CFT, and the related brane-world models.

Consider, for example, the AdS$_5\times S^5$ compactification of type
IIB theory, which gives $D=5$, $N=8$ gauged supergravity.  The
isometries of $S^5$ lead to a $SO(6)$ gauging, which in turn may be
identified with the $SO(6)$ $R$-symmetry of the $D=4$, $N=4$
super-Yang-Mills theory on the boundary.  The AdS/CFT duality then
indicates that $R$-charged black holes would couple to corresponding
$R$-charged operators of the $N=4$ SYM theory.  Such black holes, both
supersymmetric \cite{bcs1} and non-supersymmetric (non-extremal)
\cite{bcs2}, have recently been constructed in gauged $N=2$ supergravity,
which is a consistent truncation of the $N=8$ theory.  While the abelian
truncation of $N=8$ supergravity yields exactly three vector multiplets
(corresponding to the STU model), the solutions were in fact obtained for
the $N=2$ theory coupled to an arbitrary number of vector multiplets.

Since the supersymmetric $R$-charged black holes are expected to
satisfy a BPS-type no-force condition, one may believe that extremal
multi-center solutions would also exist.  However, unlike for their
ungauged counterparts, where multi-center black hole and $p$-brane
configurations are well explored and understood, only few results have
been obtained for multi-center solutions in gauged supergravities.  The
difficulty here lies in the fact that such $R$-charged black holes are
asymptotic to anti-de Sitter space.  The presence of the negative
cosmological constant suggests that only single-center solutions will be
static.  While time-dependence is perhaps natural in a cosmological
background, this added complication has yet to be fully considered in
constructing black hole solutions to gauged supergravities.

Previously, multi-center black holes that are asymptotic to de Sitter
space have been obtained in \cite{kt,london}.  As expected, these
solutions are time-dependent.  Nevertheless the time dependence enters
in a natural manner, and asymptotically yields the cosmological metric
\begin{equation}
\label{eq:dsm}
ds^2=-dt^2+e^{-2gt}d\vec x\cdot d\vec x,
\end{equation}
which has constant positive curvature.  On the other hand, for the
present purpose, one would actually like to obtain solutions in a
background of negative curvature, since $D=5$, $N=2$ gauged supergravity
naturally leads to an anti-de Sitter vacuum.  While such multi-center
solutions remain elusive, they may formally be obtained by sending $t\to it$
in the de Sitter metric, (\ref{eq:dsm}).  Alternatively, this Wick rotation
could be avoided by instead taking an imaginary coupling $g$ \cite{london}.
Of course, neither of these possibilities are entirely satisfactory.  But
nevertheless this approach provides a partial solution to the problem of
constructing multi-center anti-de Sitter black holes.  Furthermore, as many
investigations of the AdS/CFT conjecture have in fact been performed in
Euclidean space, it is perhaps not unreasonable after all to focus on the
Wick rotated theory (although in the present case, this Wick rotation
yields a pure imaginary gauge potential).

The multi-center solutions of \cite{kt,london} have been constructed in the
context of a cosmological Einstein/Maxwell theory.  As shown in
\cite{london}, up to an imaginary coupling $g$, this theory may be viewed as
the bosonic sector of pure $D=5$, $N=2$ supergravity.  Within this framework,
the black holes were shown to be supersymmetric through the explicit
construction of Killing spinors \cite{london}.  In this paper, we extend the
work of \cite{kt,london} to the case of $N=2$ supergravity with an arbitrary
number of vector multiplets.  We construct general supersymmetric
multi-centered black hole solutions and their corresponding Killing
spinors.  Some of these $N=2$ models are the gauged versions of those
obtained from M-theory compactified on a Calabi-Yau threefold.

Note that in five dimensions, supersymmetric black holes for the pure $N=2$
ungauged supergravity were first constructed in \cite{gkltt}. The metric
for these black holes are of the Tanghelini form \cite{frt}.
Subsequently, extreme black holes for $N=2$ ungauged supergravity
coupled to abelian vector multiplets were considered in
\cite{sabra1,sabra2,sabra3}. An important feature of these black holes is
that, for those with non-singular horizons, the entropy can be expressed
in terms of the extremum of the central charge.  Furthermore, the scalar
fields take fixed values at the horizon independent of their initial
values at spatial infinity \cite{feka}.  BPS black holes were
constructed in four-dimensional $N=2$ gauged supergravity in
\cite{Sabra:1999ux}.

%%%%%%%%%%%%%%%%%%%%

\section{$D=5$, $N=2$ Gauged Supergravity}

We start with five-dimensional $N=2$ gauged supergravity coupled to
$n-1$ abelian vector multiplets \cite{Gunaydin:1984bi,Gunaydin:1985ak}.
The fields in this theory
consists of a graviton $g_{\mu\nu}$, gravitino $\psi_\mu$, $n$ vector
potentials $A_\mu^I$ ($I=1,2,\ldots,n$), $n-1$ gauginos $\lambda_i$
and $n-1$ scalars $\phi^i$ ($i=1,2,\ldots,n-1$).
The bosonic part of the Lagrangian is given by%
\footnote{Our conventions are as follows: We use the metric
$\eta^{ab}=(-,+,+,+,+)$ and Clifford algebra
$\{{\Gamma ^{a},\Gamma ^{b}}\}=2\eta^{ab}$.  The gravitationally covariant
derivative on spinors is
$\nabla_\mu=\partial_\mu+{\frac{1}{4}}\omega_{\mu ab}\Gamma^{ab}$ where
$\omega_{\mu ab}$ is the spin connection.  Finally, antisymmetrization
is with weight one, so
$\Gamma^{{a}_{1}{a}_{2}\cdots {a_{n}}}={\frac{1}{n!}}\Gamma^{\lbrack {a_{1}}}
\Gamma^{{a_{2}}}\cdots\Gamma^{{a_{n}}]}$.}
\begin{eqnarray}
e^{-1}{\cal {L}} &=&\frac{1}{2}R+g^2V
-{\frac{1}{4}}G_{IJ}F_{\mu\nu}^{I}F^{\mu \nu J}
-\frac{1}{2}{\cal G}_{ij}\partial _{\mu }\phi^{i}\partial^{\mu}\phi^{j}
\nonumber\\
&&+{\frac{e^{-1}}{48}}\epsilon ^{\mu\nu\rho\sigma\lambda}
C_{IJK}F_{\mu\nu}^{I}F_{\rho\sigma}^{J}A_{\lambda}^{K}.
\label{action}
\end{eqnarray}
The scalar potential $V$ is given by
\begin{equation}
\label{eq:pot}
V(X)=V_{I}V_{J}\left( 6X^{I}X^{J}-{\frac{9}{2}}{\cal G}^{ij}\partial
_{i}X^{I}\partial _{j}X^{J}\right),
\end{equation}
where $X^I\equiv X^I(\phi^i)$ represent the real scalar fields, and
satisfy the condition ${\cal V}=1$ where
\begin{equation}
{\cal V}={\frac{1}{6}}C_{IJK}X^{I}X^{J}X^{K}.
\end{equation}
This homogeneous cubic polynomial ${\cal V}$ defines a ``very special
geometry'' of the $N=2$ theory.

The physical quantities in (\ref{action})
can all be expressed in terms of ${\cal V}$ according to
\begin{eqnarray}
G_{IJ}&=&-{\frac{1}{2}}\partial_{I}\partial_{J}\log{\cal V}\Big|_{{\cal V}=1},
\nonumber \\
{\cal G}_{ij}&=&\partial_{i}X^{I}\partial_{j}X^{J}G_{IJ}\Big|_{{\cal V}=1},
\end{eqnarray}
where $\partial_{i}$ and $\partial_{I}$ refer, respectively, to a partial
derivative with respect to the scalar field $\phi ^{i}$ and $X^{I}$.
Furthermore, the constants $V_I$ that arise in (\ref{eq:pot}) specify the
appropriate linear combination of the vectors that comprise the $N=2$
graviphoton, ${\cal A}_\mu = V_IA_\mu^I$.

For a Calabi-Yau compactification of M-theory, ${\cal V}$ denotes the
intersection form, and $X^{I}$ and $X_{I}\equiv\frac{1}{6}C_{IJK}X^{J}X^{K}$
correspond to the size of the two- and four-cycles of the Calabi-Yau
threefold.  Here $C_{IJK}$ are the intersection numbers of the threefold.
In this Calabi-Yau case, $n$ is given by the Hodge number $h_{(1,1)}$
and $C_{IJK}$ are the topological intersection numbers.  Some useful
relations from very special geometry are:
\begin{eqnarray}
&X^{I}X_{I}=1,\qquad
\partial _{i}X_{I}=-{\frac{2}{3}}G_{IJ}\partial_{i}X^{J},\qquad
X_{I}={\frac{2}{3}}G_{IJ}X^{J},&\nonumber\\
&X_{I}\partial_{i}X^{I}=X^{I}\partial _{i}X_{I}=0.&
\label{useful}
\end{eqnarray}

In gauged supergravity theories, the supersymmetry transformations get
modified by $g$-dependent terms.  For a bosonic background, the
variations of the gravitino and gauginos are given by \cite{bcs1,bcs2}:
\begin{eqnarray}
\delta\psi_{\mu}&=&\left[{\cal {D}}_{\mu }
+{\frac{i}{8}}X_I(\Gamma_\mu{}^{\nu\rho}-4\delta_\mu^\nu\Gamma^\rho)
F_{\nu\rho}^{I}+{\frac{1}{2}}g\Gamma_{\mu}X^IV_I\right]\epsilon,\nonumber \\
\delta\lambda_{i}&=&\left[-{\frac{1}{4}}G_{IJ}\Gamma^{\mu\nu}F_{\mu\nu}^J
+{\frac{3i}{4}}\Gamma^{\mu}\partial_{\mu}X_{I}+\frac{3i}{2}
gV_{I}\right]\partial_{i}X^{I}\epsilon.
\label{gst}
\end{eqnarray}
Here ${\cal D}_\mu$ is the fully gauge and gravitationally covariant
derivative,
\begin{equation}
{\cal D}_\mu\epsilon = \left[\nabla_\mu-{\frac{3i}{2}}gV_IA_\mu^I\right]
\epsilon.
\end{equation}
The ungauged theory is obtained in the (smooth) limit $g\to0$.

\subsection{Multi-centered solutions in the ungauged theory}

Before proceeding to the gauged theory, it is instructive to first
review the corresponding black hole solutions in the ungauged theory
\cite{sabra1,sabra2,sabra3,Gutowski:2000na}.  Using isotropic
coordinates, it was found that, for the non-rotating cases,
these solutions can be brought to the form
\begin{eqnarray}
\label{ma}
ds^{2}&=&-e^{-4U}dt^{2}+e^{2U}d\vec x\cdot d\vec x,\nonumber\\
A_{t}^I&=&e^{-2U}X^I,\nonumber\\
X_I&=&\ft13e^{-2U}H_I,
\end{eqnarray}
where $H_I$ are a set of harmonic functions,
\begin{equation}
\label{eq:uharm}
H_{I}=h_{I}+\sum_{j=1}^{N}\frac{q_{I\,j}}{|\vec x-\vec x_j|^2}.
\end{equation}
It is convenient to define the rescaled coordinates
\begin{equation}
Y_{I}=e^{2U}X_{I},\qquad Y^{I}=e^{U}X^{I},
\end{equation}
in which case the function $U$ may be written as
\begin{equation}
e^{3U} = \fft16C_{IJK}Y^IY^JY^K.
\end{equation}
Note that the constants $h_I$ are related to the values of the scalars
at infinity, while the $q_{I\,j}$ are the electric charges.

It is straightforward to verify that this is a half BPS solution.  In the
absence of gauging, the supersymmetry transformations (\ref{gst}) reduce to
\begin{eqnarray}
\delta\psi_{\mu} &=&\left[\nabla_{\mu}+{\frac{i}{8}}X_{I}
(\Gamma_{\mu}{}^{\nu\rho}-4\delta_{\mu}^{\nu}\Gamma^{\rho})F_{\nu\rho}{}^{I}
\right]\epsilon, \nonumber\\
\delta\lambda_{i}&=&\left[{\frac{3}{8}}\Gamma^{\mu\nu}F_{\mu\nu}^{I}
\partial_{i}X_{I}-{\frac{i}{2}}g_{ij}\Gamma^{\mu}\partial_{\mu}\phi^{j}
\right]\epsilon \nonumber\\
&=&\left[-{\frac{1}{4}}G_{IJ}\Gamma^{\mu\nu}F_{\mu\nu}^{J}+{\frac{3i}{4}}
\Gamma^{\mu}\partial_{\mu}X_{I}\right] \partial_{i}X^{I}\epsilon.
\end{eqnarray}
As usual, the resulting Killing spinor equations are solved by balancing
the gauge fields with the metric for $\delta\psi_\mu$, and the gauge
fields with the scalars for $\delta\lambda_i$.  The resulting Killing
spinors are given by $\epsilon=e^{-U}\epsilon_{0}$ where $\epsilon_{0}$ is
a constant spinor satisfying $\Gamma_{0}\epsilon_{0}=i\epsilon_{0}$.

In the absence of scalar fields, {\it i.e.}~for pure $N=2$ supergravity
theory, this solution reduces to \cite{gkltt}
\begin{eqnarray}
ds^{2} &=&-H^{-2}dt^{2}+Hd\vec{x}\cdot d\vec{x}, \nonumber\\
A_{t} &=&3H^{-1}.
\end{eqnarray}
For a single center black hole, the harmonic function may be written as
$H=1+\fft{q}{r^2}$, so that
\begin{equation}
ds^2=-\left(1+\fft{q}{r^2}\right)^{-2}dt^2
+\left(1+\fft{q}{r^2}\right)(dr^2+r^2d\Omega_3^2).
\end{equation}
This metric may be rewritten in a Schwarzschild form by defining a new
radial coordinate $\rho^2 = r^2+q$, in which case the solution becomes
\begin{eqnarray}
ds^2&=&-\biggl(1-\fft{q}{\rho^2}\biggr)^2 dt^2
+\biggl(1-\fft{q}{\rho^2}\biggr)^{-2}d\rho^2 +\rho^2d\Omega_3^2,\nonumber\\
F_{t\rho}&=&-3\partial_\rho\biggl(1-\fft{q}{\rho^2}\biggr).
\end{eqnarray}

\subsection{Single-centered solutions in the gauged theory}

Turning on the gauging now introduces a scalar potential which has the
effect of generating a negative cosmological constant.  Spherically
symmetric BPS electric solutions were found for this case in
\cite{bcs1,bcs2} and are given by
\begin{eqnarray}
\label{gs}
ds^{2} &=&-e^{-4U}f dt^{2}+e^{2U}(f^{-1}dr^{2}+r^{2}d\Omega_3^2),\nonumber\\
A_{t}^{I} &=&e^{-2U}X^{I},\nonumber\\
X_{I} &=&\ft13 e^{-2U} H_{I},
\end{eqnarray}
where
\begin{equation}
\label{eq:fsol}
f = 1+g^{2}r^{2}e^{6U}.
\end{equation}
These solutions are similar to those of the ungauged theory, (\ref{ma}),
with the only modification being the introduction of the function $f$
which depends on the square of the coupling constant $g$; for $g=0$,
we recover the solutions of the ungauged case.  The form of this metric,
with the function $f$, is identical to that used in ``blackening'' the
standard $p$-brane solutions \cite{Duff:1996hp}.  This was in fact
used to advantage in \cite{bcs2,Duff:1999gh} to blacken
these AdS solutions.

The solutions (\ref{gs}) admit Killing
spinors satisfying the projection condition \cite{bcs1}
\begin{equation}
\epsilon=-f^{-1/2}\Big(i\Gamma _{0}+gre^{3U}\Gamma _{1}\Big)\epsilon.
\label{projector}
\end{equation}
As a result, these black holes preserve half of the supersymmetry of the
$N=2$ theory.  While for $g=0$ this condition reduces to the corresponding
one in the ungauged theory, for $g\ne 0$ it has a somewhat unusual
form, at least from a $p$-brane point of view.  More details and the
solution of the Killing spinor equations, (\ref{gst}), are given in
\cite{bcs1,bcs2} (see also \cite{Romans:1992nq}).

For black $p$-branes, the absence of a no-force condition precludes the
existence of static multi-center solutions.  Along these same lines, this
spherically symmetric ansatz, and in particular the explicit use of the
radial coordinate in (\ref{eq:fsol}), prevents the above solution from
being generalized to the multi-center case.  However, unlike for the case
of black $p$-branes, here the solutions we seek are still BPS.  Although
such solutions would not be static due the cosmological background,
supersymmetry ought to provide a useful guide in their construction.

\section{Multi-centered solutions in the gauged theory}

The difficulty in obtaining multi-center solutions within the above
framework may be made evident by turning off the electric charges,
whereupon the solution, (\ref{gs}), reduces to the AdS vacuum written
in the form
\begin{equation}
ds^2=-(1+g^2r^2)dt^2+{dr^2\over 1+g^2r^2}+r^2d\Omega_3^2.
\end{equation}
This form of the AdS metric singles out a preferred radial coordinate,
and hence is ill suited as a starting point for the construction of
multi-centered solutions.  Instead, it would be more natural to use
an isotropic metric ansatz of the form
\begin{equation}
\label{eq:metans}
ds^{2}=-e^{2A}dt^{2}+e^{2B}d\vec x\cdot d\vec x,
\end{equation}
where $A$ and $B$ are functions of $(t, \vec x)$.

Before turning to the full construction of the solution, it is
instructive to examine the form of the AdS vacuum given this isotropic
ansatz.  Unfortunately it is at this stage that one encounters the
difficulty of confronting either imaginary time or coupling.  For a
multi-center solution, we insist on starting with a space-isotropic%
\footnote{If this condition were relaxed, one could select a `preferred'
space coordinate, $z$, and write the metric in Poincare form, $A=B=-\log
gz$.  However this would not lead to a suitable background from which to
construct {\it localized} black holes.}
AdS vacuum metric, so that $A$ and $B$ can only be functions of time.  In
this case, $A(t)$ may be transformed away, leaving only $B(t)$ to
consider.  The natural choice $B=-2gt$, given in (\ref{eq:dsm}),
however yields a constant positive curvature metric
\begin{equation}
R_{\mu\nu\rho\sigma}=g^2(g_{\mu\rho}g_{\nu\sigma}-g_{\mu\sigma}g_{\nu\rho}).
\end{equation}
Thus one is left with little choice but to take $gt \to igt$, yielding
\begin{equation}
\label{eq:vacads}
ds^{2}=-dt^{2}+e^{-2igt}d\vec x\cdot d\vec x.
\end{equation}
To obtain a real metric, one may either performing a Wick rotation%
\footnote{Note, however, that in Wick rotating the solution, the gauge
potential $A_t^I$ becomes pure imaginary.  Thus in either case one ends
up with imaginary or complex quantities in the gauge sector.},
$t\to it$, or transform to imaginary coupling $g\to ig$.

\subsection{Construction of the solutions}

With this AdS vacuum in mind, our aim is to find supersymmetric
multi-black hole solutions through examination of the Killing spinor
equations.  While these first order equations are often more convenient
to work with, the complete solution must still be fixed through the use
of at least one equation of motion.  The convenient choice here is to
use the Bianchi identities and equations of motion of the gauge fields
of the theory.

We start with the supersymmetry transformation of the gravitino.  For
a bosonic background and the time-dependent ansatz, this becomes
\begin{eqnarray}
\delta \psi_{t} &=&\left[\partial_{t}-\ft{3i}{2}gV_{I}A_{t}^{I}
+\ft12 g e^AX^{I}V_{I}\Gamma_0\right]\epsilon
-\ft12\Gamma^m\left[\partial_me^A\Gamma_0+iX_{I}F_{tm}{}^{I}\right]
\epsilon,\nonumber\\
\delta\psi_m&=&\left[\partial_m-\ft{i}2e^{-A}X_IF_{tm}{}^I\Gamma_0\right]
\epsilon
+\ft14\Gamma_m{}^n\left[2\partial_nB+ie^{-A}X_IF_{tn}{}^I\Gamma_0\right]
\epsilon\nonumber\\
&&\qquad-\ft12\Gamma_m\left[e^{-A}\partial_tB\Gamma_0-gX^IV_I\right]\epsilon.
\end{eqnarray}
Note that the corresponding transformation in the ungauged limit is
restored by setting $g=0$ and choosing all quantities to be time
independent.  The conditions for obtaining a vanishing transformation
may then be grouped into ones common to the ungauged limit:
\begin{eqnarray}
\label{eq:gtouga}
\left[e^A\partial_mA-iX_IF_{tm}{}^I\Gamma_0\right]\epsilon&=&0,\nonumber\\
\left[e^A\partial_m(-2B)-iX_IF_{tm}{}^I\Gamma_0\right]\epsilon&=&0,\nonumber\\
\left[2e^A\partial_m-iX_IF_{tm}{}^I\Gamma_0\right]\epsilon&=&0
\end{eqnarray}
and new conditions specific to the gauged supergravity theory:
\begin{eqnarray}
\label{eq:gtonew}
\left[2\partial_t-3igV_IA_t^I+ge^AX^IV_I\Gamma_0\right]\epsilon
&=&0,\nonumber\\
\left[\partial_tB+ge^AX^IV_I\Gamma_0\right]\epsilon&=&0.
\end{eqnarray}

In parallel with the ungauged theory, the vanishing of the first set of
equations, (\ref{eq:gtouga}), suggests the imposition of the condition
$\Gamma_{0}\epsilon =i\epsilon$ on the Killing spinors.  This leads to
the requirement
\begin{equation}
X_{I}F_{tm}{}^{I}=-\partial_{m}e^A,
\label{firsts}
\end{equation}
as well as a relation between the metric functions $A$ and $B$,
\begin{equation}
\label{eq:ABrel}
B(t,\vec x)=-\ft12A(t,\vec x)+f(t).
\end{equation}
In addition, the Killing spinor must satisfy
\begin{equation}
\label{eq:kspre}
\epsilon(t,\vec x)=e^{\fft12A(t,\vec x)}\hat\epsilon(t).
\end{equation}
We will complete the construction of the Killing spinors and return to
the additional conditions, (\ref{eq:gtonew}), below.

For now, we turn to the gaugino transformation, given by
\begin{eqnarray}
\delta\lambda_i&=&-\ft{3i}4\left[e^{-A}\partial_tX_I\Gamma_0-2gV_I\right]
\partial_iX^I\epsilon\nonumber\\
&&-\ft14\Gamma^m\left[2e^{-A}G_{IJ}F_{tm}{}^J\Gamma_0-3i\partial_mX_I\right]
\partial_iX^I\epsilon.
\end{eqnarray}
Thus two equations are obtained:
\begin{eqnarray}
\label{eq:ggo}
\left[2e^{-A}G_{IJ}F_{tm}^{J}+3i\partial_{m}X_{I}\Gamma_0\right]
\partial_{i}X^{I}\epsilon &=&0,\\
\label{eq:ggnew}
\left[\partial_{t}X_{I}+2ge^AV_{I}\Gamma_0\right]
\partial_{i}X^{I}\epsilon &=&0.
\end{eqnarray}
The first equation, also present in the ungauged theory, together with
(\ref{firsts}) implies the following connection:
\begin{equation}
\label{eq:fxrel}
{G}_{IJ}F_{tm}^{J}=\frac{3}{2}e^{2A}\partial_{m}(e^{-A}X_{I})
\end{equation}
This can be verified by using the relations (\ref{useful}).
In particular, combining (\ref{eq:fxrel}) with (\ref{useful}), we obtain
\begin{eqnarray}
e^{-A}{G}_{IJ}F_{tm}{}^{J}\partial_{i}X^{I}
&=&\ft{3}{2}e^{A}\partial_{m}(e^{-A}X_{I})\partial_{i}X^{I}
=\ft{3}{2}\partial_{m}X_{I}\partial_{i}X^{I},\nonumber\\
X_{I}F_{tm}{}^{I}=\ft{2}{3}G_{IJ}X^{J}F_{tm}^{I}
&=&e^{2A}\partial_{m}(e^{-A}X_{J})X^{J}=-\partial_{m}e^{A},
\end{eqnarray}
which solves both (\ref{firsts}) and (\ref{eq:ggo}).
Therefore one can easily deduce that the gauge fields and scalars are
related according to
\begin{equation}
A_{t}^{I}=e^{A}X^{I}.
\label{gp}
\end{equation}
To no surprise, this relation is identical to that found previously for
the multi-center solution in the ungauged theory, (\ref{ma}), and for
the single-center solution in the gauged theory, (\ref{gs}).

Until now we have only used supersymmetry to fix the relation between
the bosonic fields of the solution.  To proceed, we turn to the
equation of motion for the gauge fields:
\begin{equation}
\nabla_\nu(G_{IJ}F^{\mu\nu\,J})=\ft1{16}C_{IJK}
\epsilon^{\mu\nu\lambda\rho\sigma}F_{\nu\lambda}^{J}F_{\rho\sigma}^{K}.
\end{equation}
For a purely electric solution, the right hand side of the equation
vanishes.  There are two cases, corresponding to $\mu=t$ and $\mu=n$.
Inserting the metric ansatz, (\ref{eq:metans}), and making use of the
relations (\ref{eq:ABrel}) and (\ref{gp}), the gauge equation of motion
then yields
\begin{eqnarray}
\partial_{m}\partial_m(e^{-A+2f(t)}X_I) &=&0, \nonumber\\
\partial_{m}\partial_t(e^{-A+2f(t)}X_I) &=&0.
\end{eqnarray}
The first equation suggests a solution in terms of a harmonic function
\begin{equation}
e^{-A}X_{I}=\frac{1}{3}H_{I}(t,\vec x),
\end{equation}
while the second equation specifies the time dependence of $H_I$.  The
factor of 1/3 proves to be a convenient normalization \cite{bcs1}.
As shown in \cite{kt,london}, the solution is given simply by choosing
\begin{equation}
\label{eq:gharm}
H_I\equiv H_I(e^{f(t)}\vec x)=h_I +
\sum_{j=1}^{N}\frac{q_{I\,j}}{e^{2f(t)}|\vec x-\vec x_j|^2}.
\end{equation}

At this stage, we have satisfied all Killing spinor equations common
to both the gauged and ungauged cases.  What remains to be worked out
are the new equations specific to the gauged theory, (\ref{eq:gtonew})
for the gravitino and (\ref{eq:ggnew}) for the gaugino.  The function
$f(t)$ can be obtained from the vanishing of the gaugino transformation,
which gives
\begin{equation}
\left[\partial_tH_I+6igV_I\right]\partial_iX^I\epsilon=0,
\end{equation}
where we have used the special geometry relation $X_{I}\partial_iX^I=0$
to eliminate a term proportional to the time derivative of $e^A$.
In fact, this same relation allows the above equation to be satisfied
provided the quantity in the brackets is proportional to
$X_{I}$ and thus $H_{I}$.  Since
\begin{equation}
\label{eq:htrel}
\partial_tH_I=-2(H_I-h_I)\partial_tf,
\end{equation}
we find the solution $f(t)=-igt$, where we have used $V_I=h_I/3$.  This
imaginary time solution yields a metric
\begin{equation}
ds^{2}=-e^{2A}dt^{2}+e^{-A}e^{-2igt}d\vec{x}\cdot d\vec{x},
\end{equation}
which is asymptotic to the vacuum AdS metric of (\ref{eq:vacads}).
The metric function $A$ may be determined from the underlying very
special geometry, which implies that
\begin{equation}
e^{-\fft32A}={\cal V}(Y)={\frac{1}{6}}C_{IJK}Y^{I}Y^{J}Y^{K}.
\end{equation}

\subsection{Killing spinors}

We now return to the additional gravitino variations, (\ref{eq:gtonew}),
and solve for the Killing spinors.  Using the solution for the gauge fields,
(\ref {gp}), we obtain the following two equations for the Killing spinor:
\begin{eqnarray}
\label{kara}
\left[{\partial}_{t}(-B)-{i}ge^{A}X^{I}V_{I}\right] \epsilon=0,\nonumber\\
\left[{\partial}_{t}-{i}ge^{A}X^{I}V_{I}\right] \epsilon=0.
\end{eqnarray}
Combining these equations, we see that the Killing spinor must have the form
\begin{equation}
\epsilon(t,\vec x)=e^{-B(t,\vec x)}\tilde\epsilon(\vec x)
=e^{\fft12A(t,\vec x)}e^{-f(t)}\tilde\epsilon(\vec x),
\end{equation}
which is only consistent with the previous condition, (\ref{eq:kspre}), for
constant $\tilde\epsilon(\vec x)$, {\it i.e.}~$\tilde\epsilon(\vec
x)=\epsilon_0$ and $\hat\epsilon(t)=e^{-f(t)}\epsilon_0$.

In order to check whether the first equation of (\ref{kara}) is satisfied
we evaluate the quantity $gX^{I}V_{I}$. This can be done using our ansatz and
the relations of very special geometry. First, using our solution, we write
\begin{eqnarray}
gX^{I}V_{I} &=&\fft13gX^{I}(H_{I}-(H_I-h_I))\nonumber\\
&=&ge^{-A}-\fft12\fft{g}{\partial_t f}\partial_te^{-A},
\end{eqnarray}
where we have used $e^{-A}=\fft13X^IH_I$ as well as the relation
(\ref{eq:htrel}) for the time derivative of $H_I$.  From this, we
obtain
\begin{eqnarray}
gX^{I}V_{I}=\fft{g}{\partial_tf}e^{-A}\partial_t(f-\ft12A)
=\fft{g}{\partial_tf}e^{-A}\partial_tB,
\end{eqnarray}
which indeed satisfies (\ref{kara}) since $f(t)=-igt$.  As a result, the
Killing spinor has the simple expression
\begin{equation}
\label{eq:ksf}
\epsilon =e^{-B}\epsilon_{0}=e^{\ft12A}e^{igt}\epsilon_{0}.
\end{equation}
For $g=0$, this expression for the Killing spinor, as well as the entire
solution, reduces to that of the ungauged theory.  Furthermore, the periodic
time coordinate in AdS is apparent in (\ref{eq:ksf}), with
$t\to t+2\pi/g$.

%%%%%%%%%%%%%%%%%%%%

\section{Discussion}

To summarize, we have explored the construction of multi-center black
hole solutions to $D=5$, $N=2$ gauged supergravity.  Extending the work
of \cite{kt,london}, we have found imaginary time (or imaginary coupling)
solutions breaking exactly half of the supersymmetry.  These solutions
may be written in the form
\begin{eqnarray}
\label{eq:finsol}
ds^{2} &=&-e^{2A}dt^{2}+e^{-A}e^{-2igt}d\vec{x}\cdot d\vec{x},\nonumber\\
A_t^{I} &=&e^{A}X^{I},\nonumber\\
X_{I} &=&\ft13e^AH_{I},
\end{eqnarray}
with
\begin{equation}
e^{-\fft32A} ={\cal V}(Y)=\fft16C_{IJK}Y^{I}Y^{J}Y^{K}.
\end{equation}
In addition to the explicit factor $e^{-2igt}$ in the metric, time
dependence also arises in the harmonic functions
\begin{equation}
\label{eq:finharm}
H_I(t,\vec x)=H_I(e^{-igt}\vec x)=h_I+e^{2igt}\sum_{j=1}^N\fft{q_{I\,j}}
{|\vec x-\vec x_j|^2}.
\end{equation}
This inclusion of time dependence naturally generalizes the multi-center
solution to the ungauged theory, (\ref{ma}) and (\ref{eq:uharm}).

This solution may be made more concrete by considering the so-called $STU=1$
model ($X^{1}=S$, $X^{2}=T$, $X^{3}=U$). From (\ref{eq:finsol}), we find
\begin{equation}
\label{judy}
e^{-A}TU =H_{0},\qquad
e^{-A}SU =H_{1},\qquad
e^{-A}ST =H_{2},
\end{equation}
where $H_{0}$, $H_{1}$ and $H_{2}$ are harmonic functions of the form
(\ref{eq:finharm}). Eqn.~(\ref{judy}) together with the fact that $STU=1$
implies the following solution for the metric and the moduli fields:
\begin{equation}
e^{-3A}=H_{0}H_{1}H_{2}
\end{equation}
and
\begin{equation}
S=\left( {\frac{H_{1}H_{2}}{H_{0}^{2}}}\right) ^{\fft13},\qquad
T=\left( {\frac{H_{0}H_{2}}{H_{1}^{2}}}\right) ^{\fft13},\qquad
U=\left( {\frac{H_{0}H_{1}}{H_{2}^{2}}}\right) ^{\fft13}.
\end{equation}
The gauge fields are then given simply by
\begin{equation}
A_t^1=\fft1{H_0},\qquad
A_t^2=\fft1{H_1},\qquad
A_t^3=\fft1{H_2}.
\end{equation}

It may be noted that this construction of time-dependent
multi-center black hole solutions has a straightforward generalization
to arbitrary dimensional AdS black holes.  The most straightforward
cases would be to four and seven dimensions, where the corresponding
gauged supergravity theories have been well studied.  As a
generalization, one may also imagine seeking multi-centered $p$-brane
solutions asymptotic to AdS.  However in this case even single
$p$-branes in AdS have not yet been fully explored (but see
\cite{Chamseddine:2000xk,Klemm:2000nj} for the case of a magnetic string
in AdS$_5$).  Recently, a
procedure was given in \cite{Lu:2000xc,Cvetic:2000gj} to lift certain
multi-centered solutions in Poincar\'e supergravity to corresponding
solutions to gauged supergravities.  Starting from a multi-centered
$(p-1)$-brane in $(d-1)$ dimensions, this configuration may be lifted to
yield a multi $p$-brane solution in AdS$_d$, with a resulting
metric in ``horospheric'' form, {\it i.e.}~with a metric factor
$e^{-2gz}$ instead of $e^{-2igt}$ where $z$ is the (spatial) lifting
coordinate.  In the present case, we have little choice but to use time
in this construction, as it is the only coordinate longitudinal to a
black hole solution.  This suggests that the multi-center solutions we
have found may be viewed as lifted instantons of a $(d-1)$-dimensional
theory.  It is because these are instantons that one ends up with an
Euclideanized version of the theory.  Finally, it remains an open issue
to construct true multi-centered solutions of the gauged supergravity
theory without resorting to any Wick rotation or use of an imaginary
coupling.

\bigskip

\section*{Acknowledgments}
We wish to thank F.~Dowker for clarifying the issue of having the gauge
potential become imaginary in the Wick rotated theory.
JTL wishes to thank the Center for Advanced Mathematical Sciences
(CAMS) at the American University of Beirut, where part of this work
was performed.  This research was supported in part by DOE Grant
DE-FG02-95ER40899 Task G.

%%%%%%%%%%%%%%%%%%%%

\end{document}